\documentclass[12pt]{article}
\usepackage{graphicx}
\usepackage{amssymb}

\def\vv{\vspace{-1ex}}

\begin{document}

\begin{center}
{\Large\bf The 10\,Tesla $\mu$SR instrument: detector solutions}

\vspace{3ex}
A.~Stoykov, R.~Scheuermann, K.~Sedlak, J.~Rodriguez, U.~Greuter, A.~Amato

\vspace{2ex}
Paul Scherrer Institut, CH-5232 Villigen PSI, Switzerland

\end{center}

\vspace{2ex}
\noindent
Solutions to the detector system of the High-Field $\mu$SR instrument at
the Paul Scherrer Institut (PSI) in Switzerland are presented.
The strict technical requirements are fulfilled
through the application of Geiger-mode Avalanche Photodiodes.

\vspace{3ex}

\section{Introduction}
The High-Field $\mu$SR instrument is a highly challenging project
under realization at the Swiss Muon Source \cite{LMU}
of the Paul Scherrer Institut (PSI, Switzerland).
The detector system of the new spectrometer
has to satisfy strict requirements on the time resolution and compactness.
Muon-spin precession signals with frequencies of up to 1.3\,GHz
in magnetic fields up to 9.5\,T have to be detected,
the reduction of the signal amplitude should not exceed 50\,\%.
This requires an accuracy of better than 140\,ps (sigma) in measuring
the muon-positron time correlations --
a time resolution unprecedented for such high fields.
The small spiraling radius of the decay positrons in high fields ($\sim$\,1\,cm in 9.5\,T)
sets restrictions on the maximal radial dimension of the detector.
Preservation of the 10\,ppm uniformity of the magnetic field at the sample position
requires all detector components located in the vicinity of the sample to be non-magnetic.

R\&D work on the detector development for the High-Field project at PSI
has started in 2004.
It was realized that the required time resolution can hardly be achieved within
the ``standard" detector technology using photomultiplier tubes (PMTs),
the limiting factors being attenuation and broadening of the light pulses in the
indispensable light guides. Other potentially promising photosensors have been evaluated
and the choice was made in favor of Geiger-mode Avalanche Photodiodes (G-APDs) \cite{G-APD}.
These novel solid-state photodetectors deliver performance similar to that of PMTs,
being at the same time insensitive to magnetic fields, compact, and non-magnetic
(when choosing an appropriate packaging).
The potential of the G-APD based detector technology for $\mu$SR and its reliability
have been proven in \cite{BPM_NIMA2005,Stoykov_PhysB404-ALC,Stoykov_PhysB404-HMF}.
The found technical solutions and the gained experience
constituted an essential ground for working out the concept of the detector system
of the High-Field $\mu$SR instrument.

\section{High-Field $\mu$SR instrument: detector concept}
Figure~1(left) outlines the envisaged layout of the detector system of
the High-Field $\mu$SR instrument at PSI.
The detector is divided into two subsystems of Timing and Veto detectors.
The Timing detector consists of one muon counter Mt and 16 positron counters Pt
arranged in two rings for backward and forward positron detection.
These counters are located outside the cryostat and operating at ambient temperature.
Inside the cryostat, in the vicinity of the sample, the muon veto Mv
and positron validation Pv detectors, constituting the Veto subsystem, are located.
The scintillators of the Veto counters stay at a cryogenic temperature,
the scintillation light is collected by wavelength-shifting (WLS) fibers
and transported outside the cryostat to photosensors operating at room temperature.

\begin{figure}[tb]
\centering
\includegraphics[width=0.58\columnwidth,clip]{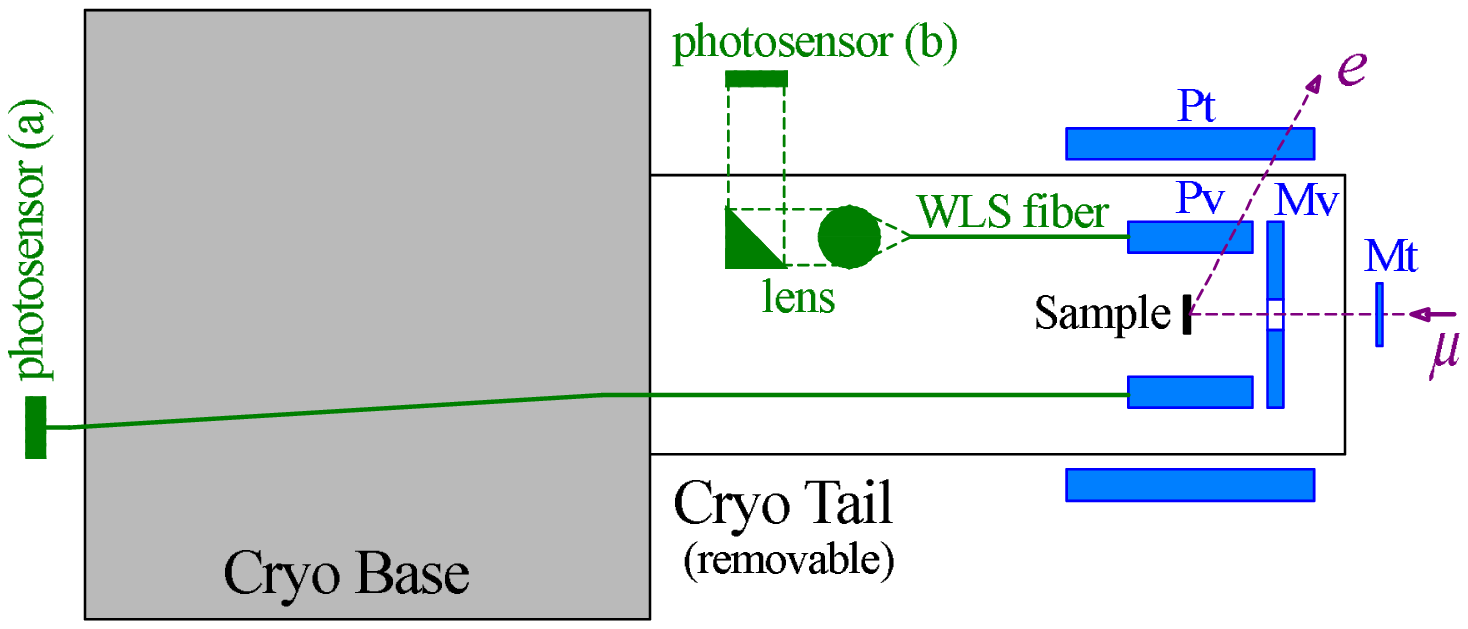} \hspace{2ex}
\includegraphics[width=0.38\columnwidth,clip]{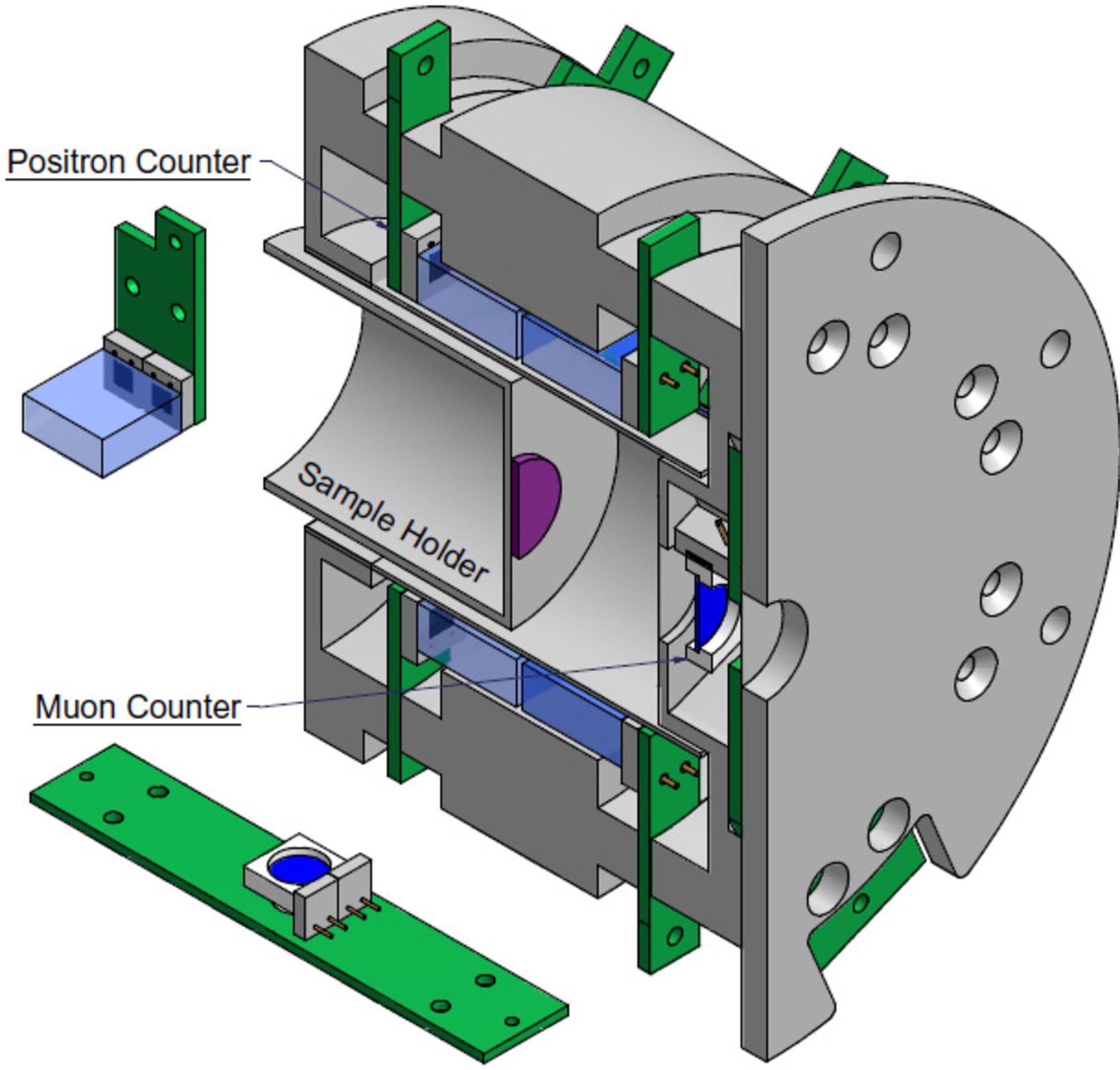}
\caption{
(Left) The sample environment and the detector layout
of the High-Field $\mu$SR instrument.
(Right)~Prototype of the Timing detector module: design view.
Plastic scintillators of type EJ-232 from Eljen are used.
Their dimensions: $\oslash 7$\,x\,0.3~mm (muon counter),
12\,x\,12\,x\,5~mm (positron counter).
Each scintillator is readout by a pair of G-APDs
of type MPPC S10362-33-050C from Hamamatsu Photonics.
}
\end{figure}

The spectrometer will be equipped with a He-flow cryostat and a dilution refrigerator.
Each cryostat will have its own detector system.
The two realizations of the Veto detector will differ in the way
how the scintillation light is transported outside the cryostat.
The construction of the flow cryostat gives sufficient freedom for
a straightforward solution -- going with WLS fibers from the scintillator to
the photosensor in a continuous (uninterrupted) manner.
In the case of dilution refrigerator realization of such approach is practically
impossible as it interferes with the foreseen cryostat design.
For this application a ``lens light guide" was developed \cite{LLG1}
allowing the light to be transported in a discontinuous manner
through the shields in the tail part of the cryostat,
thus requiring minimum adaptation of the cryostat design to integrate the detector.

The timing detectors to be used with the two cryostats are similar in design
and differ mainly in the diameter of the circle circumscribed by Pt counters --
33\,mm or 45\,mm to accommodate the flow cryostat or the dilution refrigerator, respectively.
A prototype of the timing detector (Pt-ring diameter 31\,mm)
has been build and tested in magnetic fields up to 4.8\,T.
Its design and performance are discussed below.

\section{Timing detector}

\subsection{Detector design and experimental conditions}
Figure~1(right) gives the details on the construction of the Timing detector.
Each individual counter consist of a plastic scintillator,
a photosensor constructed as a pair of G-APDs operated in series connection,
a printed board used for mounting purposes,
and an amplifier (not shown here) connected to the board via
a single $\sim 20$\,cm long cable,
supplying to the photosensor the bias voltage and taking out its signal.

The $\mu$SR measurements were performed either
on quartz (15\,x\,15\,x\,5\,mm synthetic crystal)
or silver ($\oslash 10$\,x\,1\,mm) samples.
A 29~MeV/c positive muon beam was collimated to a diameter of 5\,mm by a lead collimator
in front of the detector module.
The muon spin was rotated by $\sim 42$~degrees with respect to its initial direction
using a spin-rotator.
The external magnetic field was directed along the muon beam and the axis
of the detector module.
The analog signals from the counters were processed by constant fraction discriminators
of type PSI CFD-950 \cite{CFD950}.
Their arrival times were measured using a V1190B multihit TDC from CAEN.

\subsection{Measurements}
Figure~2 shows a $\mu$SR spectrum measured on an Ag-sample in a 4.8\,T field.
The amplitude of the muon-spin precession signal (asymmetry)
is rather large, taking into account initial rotation of the muon-spin of only 42 degrees,
and constitutes about $85 - 90$\,\% of its low-field value.
The change of the asymmetry with the magnetic field
represents a combined effect of the following factors:
1)~finite time resolution of the detector;
2)~dephasing of the muon beam;
3)~field-dependent variation of the detector acceptance.
The contribution of the time resolution to the asymmetry change
in these measurements is comparable to that from the other factors
and can not be disentangled from them.

\begin{figure}[htb]
\centering
\vspace{1ex}
\includegraphics[width=0.55\columnwidth,clip]{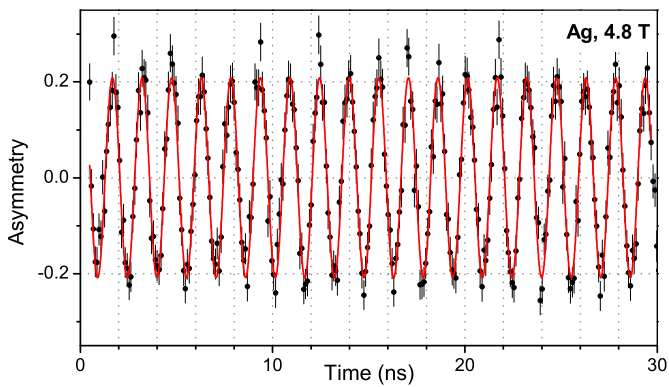}\\
\vspace*{-1ex}
\caption{
Muon spin precession measured on an Ag-sample in 4.8\,T.
}
\end{figure}

To single out the effect of the time resolution and to determine this parameter
unambiguously, measurements on the quartz sample were performed (in 0.07, 0.1, and 4.8\,T).
Fourier spectra in 0.1\,T and 4.8\,T are shown in Fig.~3(top).
The observed frequencies correspond to the precession of muons in diamagnetic states
(14\,\% fraction) and muonium (86\,\% fraction).
The amplitude of each muonium signal
is determined both by the polarization associated with the corresponding transition
and by its frequency (due to the finite time resolution).
For example, the two signals at 1560~MHz and 2394~MHz observed in 4.8~T
belong to the muonium triplett precessions
with the same polarization $P_{\rm 12} = P_{\rm 34} = 0.5$
(the small uniaxial anisotropy present
at room temperature can be neglected):
the only reason for the signal at the higher frequency to appear smaller in amplitude is
the finite time resolution of the detector.

\begin{figure}[tb]
\centering
\vspace{-1ex}
\includegraphics[width=0.59\columnwidth,clip]{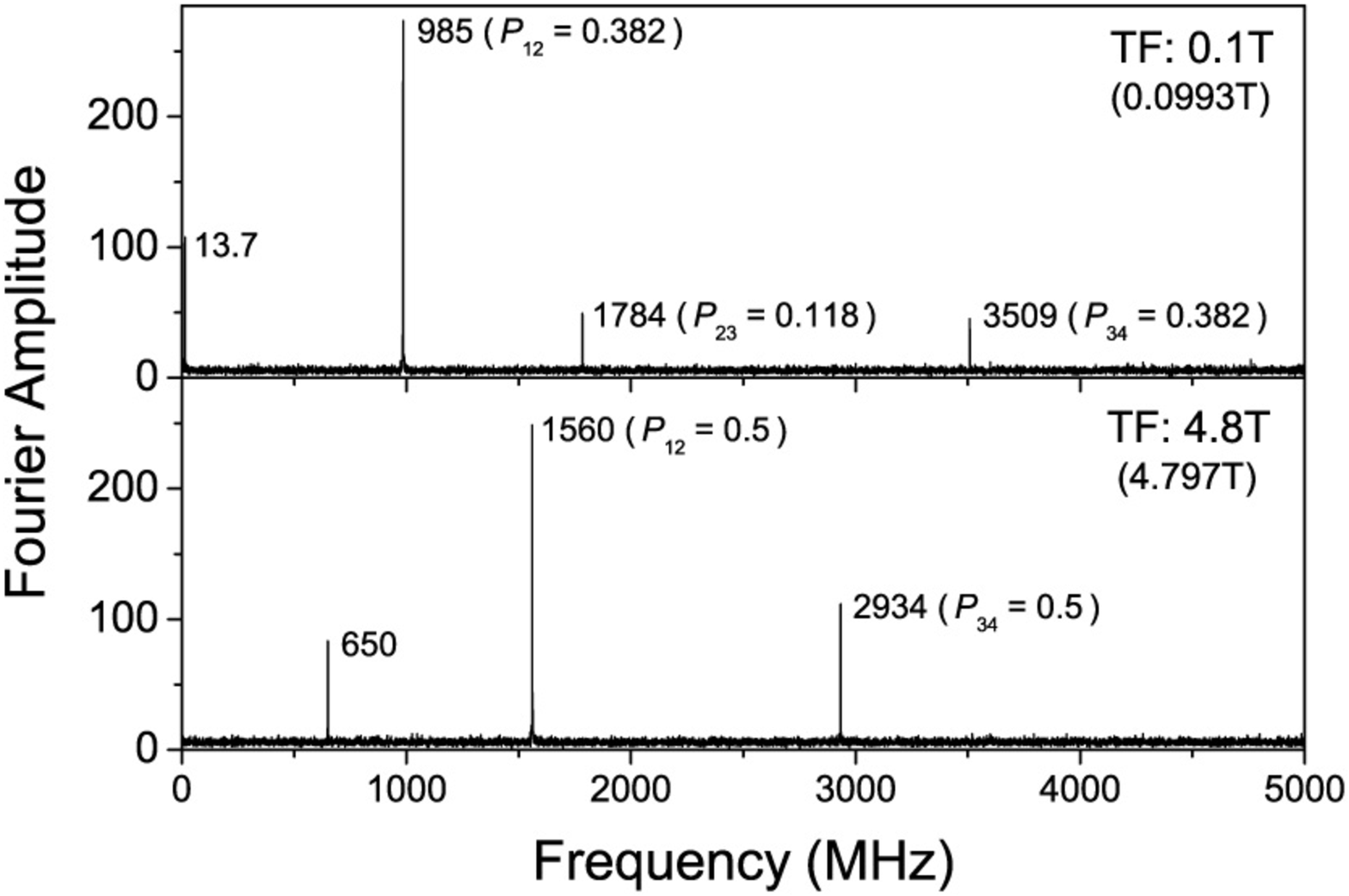} \\[1ex]
\includegraphics[width=0.59\columnwidth,clip]{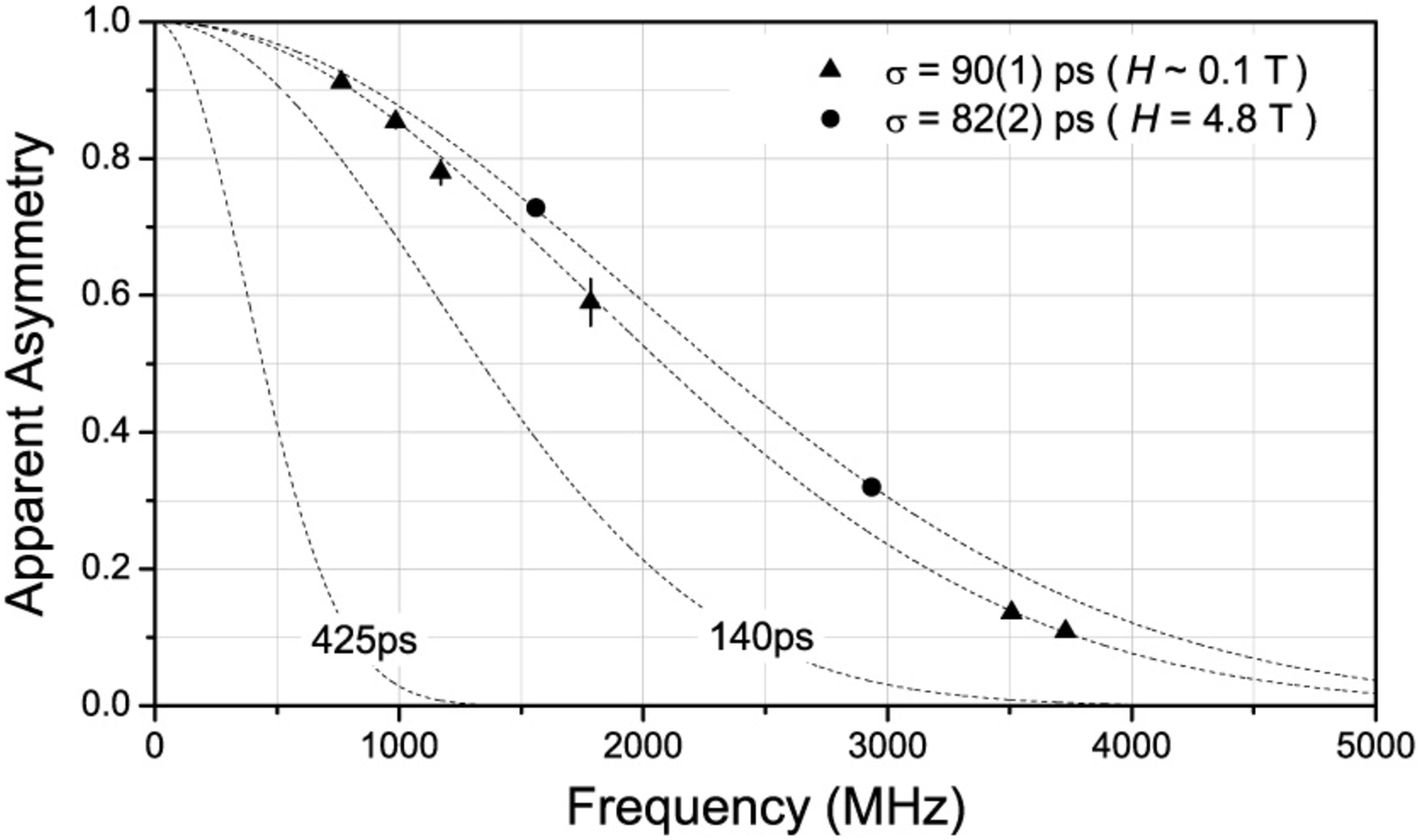}\\[-2ex]
\caption{
(Top)~Fourier spectra in quartz in transverse fields of 0.1\,T and 4.8\,T.
The isotropic part of the hyperfine coupling constant in this crystal at 290\,K
is 4495.6\,MHz.
For each muonium signal the calculated polarization $P_{\rm ij}$ is indicated.
(Bottom)~Reduction of the apparent muonium asymmetry $A_{\rm Mu}(\nu) / A_{\rm Mu} (0)$
with increasing the signal frequency measured on a synthetic quartz crystal
in transverse magnetic fields of 0.07, 0.1, and 4.8\,T.
The dashed lines represent the dependence (\ref{Eq:a(nu)}) for different values of
the time resolution $\sigma$. For comparison,
a characteristic time resolution of a ``standard" $\mu$SR instrument is $\sim 425$\,ps,
and 140\,ps is the required time resolution (upper limit) for the 9.5\,T instrument at PSI.
}
\end{figure}

For each signal of frequency $\nu$, the asymmetry obtained from the time-domain fit
divided by the corresponding polarization gives the full asymmetry
for the muonium state $A_{\rm Mu}$,
which is reduced with respect to its true value in accordance with the time resolution.
The true value of the muonium asymmetry $A_{\rm Mu}(0)$
and the time resolution $\sigma$ at the given magnetic field
were obtained by fitting the $A_{\rm Mu}(\nu)$ data
with the following function \cite{Holzschuh83}:

\begin{equation}
A_{\rm Mu}(\nu) / A_{\rm Mu}(0) = {\rm exp}[\,-\,2\,(\pi \,\sigma \, \nu)^2\,] \ .
\label{Eq:a(nu)}
\end{equation}

The low-field value of $A_{\rm Mu}(0)$ differs from that at 4.8\,T:
0.141(1) and 0.109(2), respectively.
To the difference contribute on the one hand the same factors
(apart from the time resolution),
which were responsible for the field-dependent reduction of the  asymmetry
in the Ag-sample.
On the other hand there is also a significant ``missing fraction" of the asymmetry,
which seems to increase from $\sim 25$\,\% at 0.1\,T to $\sim 35$\,\% at 4.8\,T
(the low-field Ag-asymmetry is about 0.22).
The existence of the ``missing fraction" is attributed
to delayed muonium formation \cite{Brewer_PhysB289}.

The detector time resolution, see Fig.~3(bottom),
is $90(1)$\,ps in the low fields and $82(2)$\,ps in 4.8\,T.
The contribution of the TDC to these values is not negligible
and constitutes $\sim 50$\,ps
(this combines the intrinsic time resolution of the device of $\sim 30$\,ps and
the effect of its finite bin-width of 100\,ps).
The corrected values of the time resolution are respectively 75(1)\,ps and 65(2)\,ps.
The observed improvement of the time resolution towards the high fields
is presumably due to higher energy losses of positrons crossing the counters at
more inclined trajectories.

\subsection{Time resolution: potential of the technology}
The time resolution
can be even further improved.
To demonstrate the potential of the technology,
we built a miniature $\mu$SR-setup, consisting of a muon counter,
a sample, and a positron counter, all positioned on one axis parallel to
the incoming muon beam.
In fact, the whole setup was built inside the sample
holder that was used in the measurements described above.
The muon counter has similar construction to that used before,
but the light is now collected from the two opposite sides
of the scintillator using two pairs of G-APDs.
The muon time is calculated as the mean arrival time of the two signals.
The positron counter consists of a 5\,x\,5\,x\,5\,mm scintillator read out by one G-APD.
The smaller size of the scintillator allows to reduce the contribution
to the time resolution coming from the time variations in the light collection process.
The time resolution of the new muon and positron counters was measured to be
$26 - 32$\,ps and $\sim 32$\,ps, respectively.
The expected time resolution of the setup is then $41 - 45$\,ps,
which is more than a factor of 1.5 better than in the previous measurements.
Such time resolution should allow a rather easy detection of the frequencies
in the order of 4.5\,GHz (hyperfine splitting of muonium in vacuum) and even above.
To confirm this, measurements on the synthetic quartz sample
were done in longitudinal fields up to 0.12\,T. In this measurements we used
a V1290A TDC from CAEN (bin width 25\,ps, intrinsic time resolution $\sim 30$\,ps).
Fourier spectrum of the $\mu$SR data taken in 0.12\,T is shown in Fig.~4.
The signal corresponding to the hyperfine oscillation in muonium
is clearly seen at 5631.60(3)\,MHz.

\begin{figure}[tb]
\centering
\vspace*{-1ex}
\includegraphics[width=0.9\columnwidth,clip]{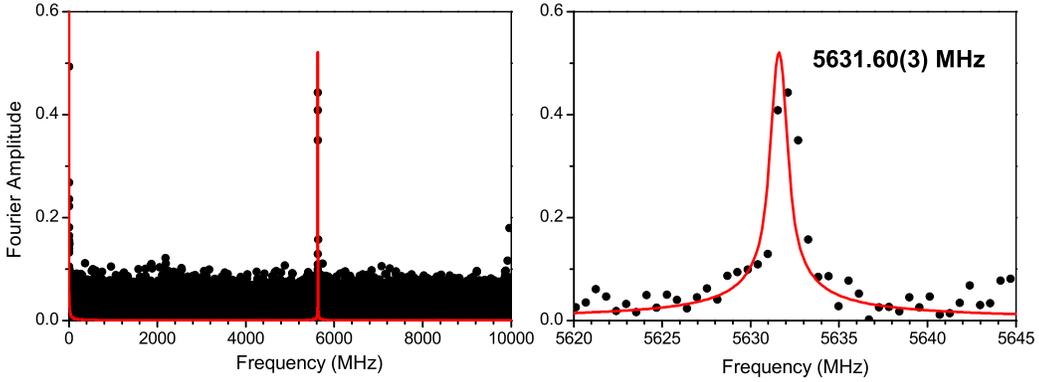} \\[-1ex]
\caption{
FFT (time window 1.75\,$\mu$s) of the hyperfine oscillation signal of muonium in
a synthetic quartz crystal in a longitudinal applied field of 0.12\,T,
measured with a dedicated high-time resolution ($\sigma \sim 45$\,ps) $\mu$SR-setup.
The red line (detailed view on the right hand side) shows the FFT of the result
of a fit of the first 1.75\,$\mu$s in the time domain,
the calculated polarization for this precession is $P_{\rm 24} = 0.32$.
}
\end{figure}

\section*{Acknowledgements}
We express our gratitude to Matthias Elender for his valuable help in designing
the detector prototypes
and in preparing the measurements. Analysis was performed using {\it musrfit} \cite{musrfit}.

\end{document}